%% file: miyabayashi_v2.tex
\documentclass[12pt]{article}
\usepackage{epsfig}

\input econfmacros.tex

\def\Title#1{\begin{center} {\Large {\bf #1} } \end{center}}

\begin{document}

\Title{Angle averages from $B$-factories}

\bigskip\bigskip


\begin{raggedright}  

{\it Kenkichi Miyabayashi\index{Miyabayashi, K.}\\
Department of Physics\\
Nara Women's University\\
KIta-Uoya-Nishi-machi, 630-8506 Nara, JAPAN}
\bigskip\bigskip
\end{raggedright}

\section{Introduction}

In the last decade, two $B$-factory experiments, the BaBar and Belle 
collaborations have been providing a lot of $CP$ violation measurements 
to perform comprehensive tests of the Kobayashi-Maskawa theory~\cite{bib_KM}. 
The three angles of the unitarity triangle, $\phi_1$, $\phi_2$ and 
$\phi_3$~\cite{footnote:phi1phi2phi3} are obtained by the relevant 
$CP$ violation measurements.
In the $B$ meson system, the complex phase in the coupling $V_{td}$ which 
contributes to $B^0-\overline{B}^0$ mixing causes the $CP$ violation 
phenomena in many cases when the neutral $B$ mesons decays into 
a $CP$ eigenstate $f_{CP}$~\cite{bib_sanda}.
Therefore the $CP$ asymmetry is to be measured as a function of 
the proper time difference ($\Delta t$) between the two $B$ meson decays 
produced by $\Upsilon(4S)$. 
The time-dependent $CP$ asymmetry $A_{CP}(\Delta t)$ is defined as
\begin{eqnarray}
A_{CP}(\Delta t) &=& 
{ {\Gamma(\overline{B}^0(\Delta t) \rightarrow f_{CP})
- \Gamma(B^0(\Delta t) \rightarrow f_{CP}) } \over
{\Gamma(\overline{B}^0(\Delta t) \rightarrow f_{CP})
+ \Gamma(B^0(\Delta t) \rightarrow f_{CP}) } }\\ \nonumber
&=& {\mathcal S}_{CP}\sin(\Delta m \Delta t) 
+ {\mathcal A}_{CP}\cos(\Delta m \Delta t), 
\end{eqnarray}
where $\Delta m$ is the mass difference between two 
mass eigenstates of the neutral $B$ meson.
The two $CP$ violation parameters, ${\mathcal S}_{CP}$ and 
${\mathcal A}_{CP}$ are
mixing-induced and direct $CP$ violation parameters, 
respectively~\cite{footnote:DCP}.
Among the three angles, $\phi_1$ and $\phi_2$ are determined by 
the time-dependent $CP$ violation measurements, while the angle $\phi_3$
is constrained by the direct $CP$ asymmetries in 
$B^{\pm} \to D K^{(*)\pm}$ decay modes. 
In this report, the most recent results of those relevant measurements 
are reviewed.
 
\section{Determination of $\phi_1 = \beta$}

For the $b \to c \bar{c}s$ transition induced decay of the neutral $B$ meson
such as $B^0 \to (c\bar{c}) K^0$, 
the Kobayashi-Maskawa theory predicts ${\cal S} = -\xi_f \sin 2 \phi_1$ and
${\cal A} =0$~\cite{bib_sanda} with very small theoretical 
uncertainty~\cite{bib_gronau},
where $\xi_f = +1$ $(-1)$ corresponds to  $CP$-even (-odd) final states and 
$\phi_1$ is one of the three interior angles of the most symmetric 
KM unitarity triangle, defined as 
$\phi_1 \equiv \mbox{arg}(-V_{cd}V^*_{cb}/V_{td}V^*_{tb})$
~\cite{footnote:phi1phi2phi3}.
Both BaBar and Belle collaborations have published their final 
measurements of $\sin 2 \phi_1$~\cite{bib_babar_sin2beta_2009,
bib_belle_sin2phi1_final}. The world average of $\sin 2 \phi_1$ is 
now $0.68 \pm 0.02$ that is a firm SM reference.
Recently the LHCb collaboration has presented their $\sin 2 \phi_1$ result and 
shown the ability to perform time-dependent $CP$ violation 
measurements~\cite{bib:LHCb_sin2phi1}.


\section{Determination of $\phi_2 = \alpha$}

The $b \rightarrow u$ transition induced $B$ decays are sensitive to 
the angle $\phi_2$. The main $B$ decay modes are $\pi\pi$, $\rho\rho$ 
and $\rho\pi$. There would be the $b \rightarrow d$ penguin contribution 
that cause complicates the extraction of the angle $\phi_2$ from the observed 
$CP$ violation, however we can solve this problem using 
an isospin analysis~\cite{bib_gandl}.
Belle has presented their measurements based on the final $\Upsilon(4S)$
data sample containing 772 M $B\overline{B}$ pairs for the time-dependent 
$CP$ violation in $B^0 \to \pi^+\pi^-$ and the search for the 
$B^0 \to \rho^0\rho^0$ decay~\cite{bib_pit_ckm2012} and BaBar has brought 
their update of $B^0 \to \rho \pi$ time-dependent Dalitz 
analysis~\cite{bib_miyashita_ckm2012}.
For the $B \to \pi\pi$ case, the isospin triangle is found to be properly 
closed, however in the $B \to \rho\rho$ case, the isospin triangle is flat, 
thus there is no plateau in the p-value plot for the $\phi_2$ determination. 
In order to have better understanding, the branching fraction measurement 
for $B^{\pm} \to \rho^{\pm} \rho^0$ decay is also important. 
As one possible attempt for this issue, Belle update of this measurement is
highly awaited. 
To have conclusive information about the possible penguin pollution in
$\rho \rho$ modes, more precise measurements of $B^0 \to \rho^0 \rho^0$ 
are necessary and Super $B$ factory statistics are required.
With the currently available statistics, $B^0 \to (\rho \pi)^0$ display
non-Gaussian behavior. At the level of 2$\sigma$, most of the $\phi_2$ region
is still allowed when considering $\rho \pi$ alone.
Except for the latest BaBar $B^0 \to (\rho \pi)^0$ result, that requires 
special care to be combined with other results because of the non-Gaussian
behavior, including all the results made available, we get 
$\phi_2 = (88.5 ^{+4.7} _{-4.4}) ^{\circ}$ from
a combined fit~\cite{bib_ckmfit}.


\section{Determination of $\phi_3 = \gamma$}

The $CP$ violation phenomena caused by the angle $\phi_3$ appear
as the direct $CP$ asymmetry in the charged $B$ meson decays into 
a neutral $D$ meson and a charged Kaon, when the neutral $D^{(*0)}$ meson
turns into the final state where both $D^{(*0)}$ and $\overline{D}^{(*0)}$ 
can decay. Recently Belle has performed the GLW~\cite{bib_GLW} measurement 
using $B^{\pm} \to D^{*0} K^{\pm}$ where $D^{*0} \to D^0 \pi^0$, 
$D^0 \to K^+K^-$ and $\pi^+\pi^-$ for $CP$-even and $D^0 \to K^0_S \pi^0$ 
and $K^0_S \eta$ for $CP$-odd final states. 
The same measurements are also carried out for $B^{\pm} \to D^{*0} K^{\pm}$, 
$D^{*0} \to D^0 \gamma$~\cite{bib_karim_ckm2012}. 
Combining with the GGSZ~\cite{bib_GGSZ} and the ADS~\cite{bib_ADS} related 
measurements, Belle's $\phi_3$ constraint is found to be 
$\phi_3 = (68 ^{+15} _{-14}) ^{\circ}$.
BaBar converted all their GLW and ADS available results into the GGSZ-familiar
cartesian variables and the angle $\phi_3$ is obtained to be 
$\phi_3 = (69 ^{+17} _{-16}) ^{\circ}$~\cite{bib_denis_ckm2012}.

\section{Summary}

Two $B$-factories have brought precise determination 
$\sin 2 \phi_1 = 0.68 \pm 0.02$ in world average.
The angle $\phi_2$ is obtained to be $(88.5 ^{+4.4} _{-4.2}) ^{\circ}$,
Belle new $B^0 \to \pi^+\pi^-$, $\rho^0 \rho^0$ results as well as 
BaBar update $B^0 \to (\rho \pi)^0$ time-dependent Dalitz analysis have 
been shown in this workshop.
Belle and BaBar have come up with each $\phi_3$ constraint, 
$(68 ^{+15} _{-14}) ^{\circ}$ by Belle and 
$(69 ^{+17} _{-16}) ^{\circ}$ by BaBar.
We still see relevant $CP$ violation measurements are active at 
$B$-factories.

\bigskip
The author's participation to the 7th International Workshop on the CKM
Unitarity Triangle (CKM2012) was supported by MEXT KAKENHI, 
Grant-in-Aid for Scientific Research on Innovative Areas, 
entitled ``Elucidation of New hadrons with a Variety of Flavors''.

\end{document}

%% file: econfmacros.tex
\def\beq{\begin{equation}}
\def\eeq#1{\label{#1}\end{equation}}
\def\eeqn{\end{equation}}

\def\beqa{\begin{eqnarray}}
\def\eeqa#1{\label{#1}\end{eqnarray}}
\def\eeqan{\end{eqnarray}}

\let\bar=\overbar

\def\Dslash{\not{\hbox{\kern-4pt $D$}}}
\def\dslash{\not{\hbox{\kern-2pt $\del$}}}

\def\msb{{\bar{\ssstyle M \kern -1pt S}}}




%% file: miyabayashi_v2.bbl
\begin{thebibliography}{99}


\bibitem{bib_KM}
M.~Kobayashi and T.~Maskawa, Prog.~Theor.~Phys. {\bf 49}, 652 (1973).

\bibitem{footnote:phi1phi2phi3}
Another naming convention $\beta(=\phi_1)$, $\alpha(=\phi_2$ and 
$\gamma(=\phi_3)$ is also used in the literatures.

\bibitem{bib_sanda}
A.~B.~Carter and A.~I.~Sanda, Phys.\ Rev.\ D \textbf{23}, 1567 (1981);
I.~I.~Bigi and A.~I.~Sanda, Nucl.\ Phys. B \textbf{193}, 85 (1981).

\bibitem{bib_gronau}
M.~Gronau, Phys.\ Rev.\ Lett.\  {\bf 63}, 1451 (1989).

\bibitem{footnote:DCP}In some cases, ${\mathcal C}_{CP} = - {\mathcal A}_{CP}$
is used to represent the direct $CP$ violation.

\bibitem{bib_babar_sin2beta_2009}
B.~Aubert {\it et al.} (BaBar Collaboration), 
Phys.\ Rev.\ D {\bf 79}, 072009 (2009).

\bibitem{bib_belle_sin2phi1_final}
I.~Adachi {\it et al.} (Belle Collaboration),
Phys.\ Rev.\ Lett.\  {\bf 108}, 171802 (2012).

\bibitem{bib:LHCb_sin2phi1}
Julian Wishahi, ``Measurement of $\Delta m_s$, $\Delta m_d$ and 
$\sin 2 \beta$ with LHCb'', talk given at the 7th International Workshop
on the CKM Unitarity Triangle (CKM2012); 
also found as LHCb-PAPER-2012-035, arXiv:1211.6093.

\bibitem{bib_gandl}
M.~Gronau and D.~London, Phys.\ Rev.\ Lett.\  {\bf 65}, 3381 (1990).

\bibitem{bib_pit_ckm2012}
Pit Vanhoefer, ``Results on $\phi_1$ and $\phi_2$ from Belle'', talk given 
at the 7th International Workshop on the CKM Unitarity Triangle (CKM2012).

\bibitem{bib_miyashita_ckm2012}
Tomo Miyashita, ``Measurement of $CP$-violating asymmetries in 
$B^0 \to (\rho \pi)^0$ using a time-dependent Dalitz plot analysis of the 
BaBar dataset'', talk given at 
the 7th International Workshop on the CKM Unitarity Triangle (CKM2012).

\bibitem{bib_ckmfit}
CKMfitter group, ``Update of $\alpha/\phi_2$ at CKM2012''.

\bibitem{bib_GLW}
M.~Gronau and D.~London, Phys. Lett. {\bf B 253}, 483 (1991);
M.~Gronau and D.~Wyler, Phys. Lett. {\bf B 265}, 172 (1991).

\bibitem{bib_karim_ckm2012}
Karim Trabelsi, ``DCPV in charmed $B$ decays and $\gamma / \phi_3$ average 
from Belle'', talk given at 
the 7th International Workshop on the CKM Unitarity Triangle (CKM2012).

\bibitem{bib_GGSZ}
A.~Giri, Y.~Grossman, A.~Soffer and J.~Zupan,
Phys. Rev. {\bf D 68}, 054018 (2003);
A.~S.~Bondar, Proceedings of Belle Dalitz Analysis meeting, 24-26 Sep. 2002.

\bibitem{bib_ADS}
D.~Atwood, I.~Dunietz and A.~Soni,
Phys. Rev. Lett. {\bf 78}, 3257 (1997); Phys. Rev. {\bf D 63}, 036005 (2001).

\bibitem{bib_denis_ckm2012}
Denis Derkach, ``Combination of $\gamma$ measurements from BaBar'', 
talk given at 
the 7th International Workshop on the CKM Unitarity Triangle (CKM2012).

\end{thebibliography}
